\documentstyle[11pt]{article}
 
\def\pa{\partial}

\def\a{\alpha} 
\def\b{\beta} 
 
\def\e{\epsilon}

\def\m{\mu}

\renewcommand{\baselinestretch}{2.0}

\begin{document}

\begin{flushright}
BRX TH--393 \\
ULB-TH96/13 \\
hep-th/9607182
\end{flushright}

\begin{center}
{\large\bf Electric-Magnetic Black Hole Duality}

S. Deser\\
{\small Department of Physics, Brandeis University, Waltham, MA 02254, USA}\\
M. Henneaux \\
{\small Facult\'e des Sciences, Universit\'e Libre de Bruxelles, Campus
Plaine C. P. 231, B-1050  Bruxelles,  \\ Belgium and \\
Centro de Estudios
Cient\'{\i}ficos de Santiago, Casilla 16443, Santiago, Chile} \\
and \\
C. Teitelboim \\
{\small Centro de Estudios
Cient\'{\i}ficos de Santiago, Casilla 16443, Santiago, Chile and\\
School of Natural Sciences, Institute for Advanced Study,
Princeton, NJ 08540, USA} \\
\end{center}

\begin{quotation}
\noindent {\bf Abstract:}

We generalize duality invariance for the free Maxwell action in 
an arbitrary background geometry to include the presence of 
electric and magnetic charges.  In particular,
it follows that the actions 
of equally charged electric and magnetic black holes are equal.
\end{quotation}

It has long been known \cite{001} that in an arbitrary background 
geometry the source-free Maxwell action is duality-invariant 
under general electric/magnetic field rotations (for a recent
review of electromagnetic duality, see \cite{Olive}).  This is the case 
despite the change of sign of the 
Maxwell Lagrangian $- \frac{1}{4} \, F^2_{\mu\nu}$ under 
the  ``duality rotation" $F_{\mu\nu} \rightarrow \,^*\!F_{\mu\nu}$.  
Due to the interest of electrically and magnetically charged 
black holes in the semiclassical regime, it is natural to study 
duality of the action there as well.  We shall show that the action, 
and not just the field equations or the stress tensor, is 
duality invariant in such a way that equivalence of the black 
hole actions and partition functions is maintained  for the charged solutions. 
In particular, we provide a simple derivation of similar recent 
conclusions of Hawking and Ross \cite{002}, 
which were reached through a somewhat laborious procedure.
As the subject of Maxwell duality 
seems always to be fraught with confusion, we will at times belabor 
the obvious to make our point clear.

When dealing with magnetically charged black holes, one 
usually determines the partition function in a sector with
fixed magnetic charge.  To make a meaningful compararison
between the electric and magnetic cases, it is therefore
necessary to compute the partition function for electrically
charged black holes also in a sector with fixed charge,
rather than fixed electric chemical potential.  For this reason, we shall 
deal in this paper exclusively with the action adapted to
the variational principle in which the charges are kept fixed. 
Technically, this is achieved by considering field histories
in the variational principle that all have the same given
electric and magnetic fluxes at infinity (see below).
Our main result is that there is a transformation of the
dynamical variables which, when combined with the corresponding
rotation of the charges, maps the action for a fixed set of
electric and magnetic charges to the action for the duality-rotated
set of fixed electric and magnetic charges. 

Consider first source-free electrodynamics in an arbitrary background
with topology $R \times \Sigma$, where the spatial sections $\Sigma$
are homeomorphic to $R^3$.  
As shown explicitly in \cite{001}, the 
Maxwell action \underline{is} invariant 
under arbitrary finite duality rotations.  In canonical form, in 
the usual 3+1 notation, it is \cite{003}
\begin{equation}%1
I_M [ {\bf E}, {\bf A}] = -\int d^4x 
[E^i\dot{A}_i + \textstyle{\frac{1}{2}} \, Ng^{-1/2} 
g_{ij} (E^iE^j +B^iB^j) - \e_{ijk}
N^iE^jB^k ]    \label{action1}
\end{equation}
where $E^i$ is the electric, $B^i \equiv \e^{ijk}\pa_j A_k$ the magnetic, 
field (both are contravariant three-densities) and all metrics are in 3-space; 
we have solved the Gauss constraint so that both  $E^i$ and $B^i$ are
identically transverse,   
$\pa_i E^i = 0 = \pa_i B^i$.  As usual, we take the field configurations
to behave asymptotically as $A_i = a_i(\theta, \phi) r^{-1} 
+ O(r^{-2})$ and $E^i = e^i(\theta, \phi) r^{-2} + O(r^{-3})$ as
$r \rightarrow \infty$.

The variation of the action under changes of
$E^i$, 
\begin{equation}
\delta_E I_M = -\int d^4x \delta E^i
(\dot{A}_i  + Ng^{-1/2}g_{ij}E^j- \e_{ijk} B^j N^k),
\label{varE}
\end{equation}
vanishes for arbitrary variations $\delta E^i$ subject
to the transversality conditions\footnote{The condition
$\delta \oint _{S^2_{\infty}} E^i dS_i =0$ is actually
a consequence of $\pa_i \delta E^i = 0$ (and
of smoothness) on spatial sections with $R^3$-topology.  We write 
it separately, however, because this is no longer
the case if $\Sigma$ has holes, as below.} $\pa_i \delta E^i = 0$
and $\delta \oint _{S^2_{\infty}} E^i dS_i =0$ if and only if the 
coefficient of $\delta E^i$ in (\ref{varE}) fulfills the condition 
\begin{equation}%3
\dot{A}_i  + Ng^{-1/2}g_{ij}E^j- \e_{ijk} B^j N^k=
\partial_i V
\label{equE}
\end{equation}
where $V$ ($\equiv A_0$) is an arbitrary function which
behaves asymptotically as $C + O(r^{-1})$:  In that
case, $\delta I_M = -\int d^4x \delta E^i
\partial_i V = - \oint _{S^2_{\infty}} \delta E^i V dS_i
= - C \delta$(electric flux) = 0.  No special conditions are 
required, on the other hand, when varying $A_i$.  Thus,
the action (1) is appropriate as
it stands, {\it i.e.}, without ``improving" it by adding
surface terms,  to the variational principle in which the
competing histories all have the same given electric
flux at infinity and thus also the same given electric
charge (here equal to zero).  

As pointed out in \cite{002},
it is necessary to allow the temporal component $V$ 
of the vector potential to approach a non-vanishing constant 
at infinity since this is what happens in the
black hole case if $V$ is required to be regular on the
horizon.  However, as we have just shown,
in order to achieve this while working with
the action (\ref{action1}), it is unnecessary to keep all
three components $E^i$ of the electric field fixed at spatial
infinity;   only  the electric flux
$\oint _{S^2_{\infty}} E^i dS_i$ must be kept constant in the
variational principle.

The action  (1) is invariant under
duality rotations.  Indeed, the finite rotation 

\renewcommand{\baselinestretch}{1}
\small\normalsize
\begin{equation}%4
\left( \begin{array}{c}
{\bf E}^\prime\\{\bf B}^\prime 
\end{array}
\right)
= R
\left( \begin{array}{c}
{\bf E}\\{\bf B} 
\end{array}
\right)
\equiv
\left( \begin{array}{ll}
\cos \theta & \sin \theta\\
-\sin \theta & \cos \theta 
\end{array}
\right)  
\left( \begin{array}{c}
{\bf E}\\{\bf B} 
\end{array}
\right)  \label{dual}
\end{equation}
\renewcommand{\baselinestretch}{2}
\small\normalsize
(which specifies the rotation of
${\bf A}$ up to a gauge transformation)
manifestly preserves all but the E\.{A} term in (1), whose separate invariance 
is also easy to check. [The surface term that one picks up at spatial infinity
from the variation of the kinetic term is easily seen to vanish with
the given asymptotic conditions].  As explained in \cite{001}, there
is no contradiction between the invariance of the
action (\ref{action1}) under (\ref{dual}) and its change of sign 
under
\begin{equation}
F_{\mu \nu} \rightarrow 
(-g)^{-\frac{1}{2}}
g_{\m\a}g_{\nu \b} \:^*\!F^{\a \b}.
\label{wrong}
\end{equation}
The point is that any transformation must be 
represented in terms of the independent field 
variables, which (5) cannot as it is most easily seen
by observing that 
$d ^*\!F$ (unlike $dF$) does
not vanish identically.  On the other hand, an explicit generator of the
rotation (4) does exist[1]\footnote{This is not always the case:
for the scalar field in 2d, for example, the duality is on
$\partial_\mu \phi \rightarrow \epsilon_{\mu \nu} \partial^\nu \phi$,
or in canonical language, on rotation of the canonical momentum to
$\phi$  and its gradient, 
but there is no generator of the equivalent of (4).  Similar difficulties
arise for $2k$-form gauge fields in $2+4k$ dimensions. While the
energy density is invariant under duality rotations, the kinetic
term is not.  Thus, duality rotations are not canonical transformations
in those cases.}.

We now turn to the black hole case and include 
electric and magnetic sources. 
To stick to the problem of interest in \cite{002}, where only the
exterior solution is considered, one can still work with
the source-free Maxwell equations but one must allow for
non-vanishing electric and magnetic fluxes at infinity.
This is possible because the spatial sections $\Sigma$ have a
hole.  There are thus two-surfaces that are
not contractible to a point, namely, the surfaces
surrounding the hole (we assume for simplicity 
a single black hole but the
analysis can straightforwardly be extended to the multi-black
hole case).

In the presence of a non-vanishing magnetic flux, the magnetic
field is given by the expression 
\begin{equation}
B^i = \e^{ijk} \partial_j A_k + B^i_S
\label{magnetic}
\end{equation}
where $B^i_S$ is a fixed field that carries the magnetic
flux,
\begin{equation}
\oint _{S^2_{\infty}} B^i_S dS_i = 4 \pi \mu,
\end{equation}
and where $B^i_T = \e^{ijk} \partial_j A_k$ is the transverse
part of $B^i$,
\begin{equation}
\partial_i B^i_T = 0, \; \oint _{S^2_{\infty}} B^i_T dS_i = 0.
\end{equation}
Following Dirac, we can take $B^i_S$ to be entirely
localized on a string running from the source-hole to infinity, 
say along the positive $z$-axis $\theta =0$.
We shall not need the explicit form of $B^i_S$ in the sequel, 
but only to remember that for a given magnetic charge $\mu$,
$B^i_S$ is completely fixed and hence is not a field to be varied 
in the action.  The only dynamical components
of the magnetic field $B^i$ are still the transverse ones, i.e.,
$A_i$.

One can also decompose the electric field as
\begin{equation}
E^i = E^i_T + E^i_L
\end{equation}
where the longitudinal part carries all the electric flux
\begin{equation}
\oint _{S^2_{\infty}} E^i_L dS_i = 4 \pi e
\end{equation}
and the transverse field obeys
\begin{equation}
\partial_i E^i_T = 0, \; \oint _{S^2_{\infty}} E^i_T dS_i = 0
\end{equation}
and can thus be written as $E^i_T = \e^{ijk} \partial_j Z_k$ for
some $Z_k$.
Given the electric charge $e$, the longitudinal electric
field is completely determined if we impose in addition, say,
that it be spherically symmetric.  As we have done above,
we shall work with a variational principle in which we have
solved Gauss's law and in which the competing histories have
a fixed electric flux $\oint _{S^2_{\infty}} E^i dS_i$ at
infinity.  This means that the longitudinal electric
field is completely frozen and that only the tranverse
components $E^i_T$ are dynamical, as for the magnetic field.

In order to discuss duality, it is convenient to treat  
the non-dynamical components of $E^i$ and
$B^i$ symmetrically.  To that end, one may either redefine $B^i_S$ by 
adding
to it an appropriate transverse part so that it shares the
spherical symmetry  of $E^i_L$, or one
may redefine $E^i_L$ by adding
to it an appropriate transverse part so that it is entirely
localized on the string.  Both choices (or, actually, any
other intermediate choice) are acceptable here.  For concreteness 
we may take the first choice; the fields then
have no string-singularity.

The Maxwell action for the problem at hand takes exactly the
same form (1), but with
$E^i$ and $B^i$ now the total electric and magnetic
fields.  Since $E^i_L$ may be taken to be time-independent
(the electric charge is constant), one may replace $E^i$ by
$E^i_T$ in the kinetic term of (1), yielding
as alternative action
\begin{equation}%12
I_M^{e,\mu} [ {\bf E}_T, {\bf A}] = -\int d^4x
[E^i_T\dot{A}_i + \textstyle{\frac{1}{2}} \, Ng^{-1/2}
g_{ij} (E^iE^j +B^iB^j) - \e_{ijk}
N^iE^jB^k ].    \label{action2bis}
\end{equation}
This amounts to dropping a total time derivative - equal
to zero for periodic boundary conditions - and
shows explicitly that the kinetic term is
purely transverse.  The
action (12) involves the  parameters $e$ and $\mu$; 
that is, one has a distinct variational
principle for each choice of $e$ and $\mu$, as we have indicated.

To discuss the surface terms that arise in the variation of the 
action, one must supplement the asymptotic behavior of the fields 
at infinity by conditions at the horizon.  In the Euclidean continuation, 
time becomes an angular variable with the horizon sitting at the 
origin of the corresponding polar coordinate system.  
Regularity then requires that 
$V \equiv A_0$ and the time derivatives 
$\dot{A}_i , \; \dot{E}^i $ all vanish at the horizon.  
We assume these conditions to be fulfilled throughout.

Consider now a duality rotation (4) acting on the dynamical 
variables $A_i$ (or $B^i_T$) and $E^i_T$.  Just as in the sourceless 
case, the kinetic term of (12) is invariant
under this transformation: it is the same kinetic term and the 
transformation law is the same;
the surface term at the horizon in the variation vanishes
because $\dot{A}_i=0$ and $\dot{Z}_i=0$ there.
Thus, if we also
rotate the (non-dynamical) components of the electric
and magnetic fields in the same way, that is, if we
relabel the external parameters $e$, $\mu$ by the 
same transformation as in (4),

\renewcommand{\baselinestretch}{1}
\small\normalsize
\begin{equation}
\left( \begin{array}{c}
e^\prime\\\mu^\prime
\end{array}
\right)
= R
\left( \begin{array}{c}
e\\\mu
\end{array}
\right) \; ,
\end{equation}
\renewcommand{\baselinestretch}{2}
\small\normalsize
the actions $I_M^{e,\mu}$ and $I_M^{e^\prime,\mu^\prime}$ are equal
since ${\bf E}$ and ${\bf B}$ enter totally symmetrically in the
energy and momentum densities.
More explicitly, if we write the longitudinal fields as 
$B^i_L = \mu V^i$, $E^i_L = eV^i$, then the relevant terms in (13) are just 
\begin{equation}%14
 -  \int d^4x \left\{ 
Ng^{-1/2} g_{ij} 
\left[ (e E^i_T  + \mu B^i_T) V^j
+ \textstyle{\frac{1}{2}} (e^2 + \mu^2 ) V^iV^j \right]
- \epsilon_{ijk} N^i V^j (e B^k_T - \mu E^k_T ) \right\} \; .
\end{equation}
For the mixed terms, it is clear that the field transformation (4) 
is just compensated by the parameter rotation (12)
(the relevant terms are just the 
``dot" and ``cross" product of the corresponding ``2-vectors"), 
while the $VV$ term is invariant under it.  To put it more formally, 
the extended duality invariance we have spelled out is
\begin{equation}
I^{e,\mu}_M [{\bf E}_T, {\bf A}_T] =
I^{e^\prime ,\mu^\prime}_M [{\bf E}_T^\prime, {\bf A}_T^\prime]\; ,
\label{invar}
\end{equation}
where the primes denote the rotated values.  To avoid any misunderstanding,
we emphasize that at no stage we have introduced a potential
for $^*\! F$.  For us,
only the usual $A_\mu$ is ever defined.

Equation (\ref{invar}) links two actions 
parameterized by different values of $(e,\mu )$. 
In particular, for the black holes without Maxwell excitations, we find
\begin{equation}%16
I_M^{e,0} [{\bf 0,0}] = I_M^{0,e}[{\bf 0,0}]
\end{equation}
as in \cite{002}.  
This equality is thus not a special artifact, but reflects a general
invariance property of the action appropriate to the
variational principle considered here, in which the electric and magnetic
fluxes are kept fixed.  The invariance of the action can
be verified along the same lines if one also includes the
dilaton field.

SD thanks the Erwin Schroedinger Institute and
the Institute for Theoretical Physics of Vienna
University and MH is grateful to LPTHE (Paris VI and Paris VII) for kind
hospitality while this work was completed.
The work of SD was supported by the National Science Foundation, 
grant \#PHY-9315811, that of MH was partly supported
by a research grant from FNRS (Belgium) and that of CT
by  institutional support to the
Centro de Estudios Cient\'{\i}ficos de Santiago provided by
SAREC (Sweden) and a group of Chilean private companies
(EMPRESAS CMPC, CGE, COPEC, MINERA LA ESCONDIDA, NOVAGAS
Transportandores de Chile, ENERSIS, BUSINESS DESIGN ASS., XEROX
Chile).


\begin{thebibliography}{99}
\bibitem{001}
S. Deser and C. Teitelboim, Phys. Rev. {\bf D13}, 1592 (1976);
S. Deser, J. Phys. {\bf A15} 1053 (1982).
\bibitem{Olive} D.I. Olive, {\em Exact Electromagnetic Duality},
hep-th/9508089.
\bibitem{002}
S.W. Hawking and S.F. Ross, Phys. Rev. {\bf D52}, 5865 (1995).
\bibitem{003}
R. Arnowitt, S. Deser, and C.W. Misner in {\it ``Gravitation, An 
Introduction to Current
Research"}, (L. Witten, ed.) (Wiley, NY, 1962). 
\end{thebibliography}
\end{document}